\documentclass[reprint,floatfix,amsmath,amssymb,aps,superscriptaddress,nofootinbib,longbibliography]{revtex4-1} 

\usepackage{graphicx}
\usepackage{float}
\usepackage{amsmath}
\usepackage{bm}

\usepackage{xcolor}
\usepackage{footnote}
\usepackage{soul} 

\begin{document}

\preprint{APS/123-QED} 

\title{Generating octave-bandwidth soliton frequency combs with compact, low-power semiconductor lasers}


\author{Travis C. Briles}
\email[]{travis.briles@nist.gov}
\affiliation{Time and Frequency Division, National Institute for Standards and Technology, Boulder, CO 80305 USA}
\affiliation{Department of Physics, University of Colorado, Boulder, CO 80309 USA}
\author{Su-Peng Yu}
\affiliation{Time and Frequency Division, National Institute for Standards and Technology, Boulder, CO 80305 USA}
\affiliation{Department of Physics, University of Colorado, Boulder, CO 80309 USA}
\author{Tara E. Drake}
\affiliation{Time and Frequency Division, National Institute for Standards and Technology, Boulder, CO 80305 USA}
\affiliation{Department of Physics, University of Colorado, Boulder, CO 80309 USA}
\author{Jordan R. Stone}
\affiliation{Time and Frequency Division, National Institute for Standards and Technology, Boulder, CO 80305 USA}
\affiliation{Department of Physics, University of Colorado, Boulder, CO 80309 USA}
\author{Scott B. Papp}
\affiliation{Time and Frequency Division, National Institute for Standards and Technology, Boulder, CO 80305 USA}
\affiliation{Department of Physics, University of Colorado, Boulder, CO 80309 USA}

\date{\today}

\begin{abstract}
We report a comprehensive study of low-power, octave-bandwidth, single-soliton microresonator frequency combs in both the 1550 nm and 1064 nm bands.  Our experiments utilize fully integrated silicon-nitride Kerr microresonators, and we demonstrate direct soliton generation with widely available distributed-Bragg-reflector lasers that provide less than 40 mW of chip-coupled laser power. We report measurements of soliton thermal dynamics and demonstrate how rapid laser-frequency control, consistent with the thermal timescale of a microresonator, facilitates stabilization of octave-bandwidth soliton combs. Moreover, since soliton combs are completely described by fundamental linear and nonlinear dynamics of the intraresonator field, we demonstrate the close connection between modeling and generation of octave-bandwidth combs. Our experiments advance the development of self-referenced frequency combs with integrated-photonics technology, and comb-laser sources with tens of terahertz pulse bandwidth across the near-infrared.

\begin{description}
\item[PACS numbers]
\pacs{1} 42.82.-m
\pacs{2} 42.60.Da
\pacs{3} 42.62.Eh
\pacs{4} 42.65.Tg

\end{description}
\end{abstract}

\maketitle

\setlength{\parskip}{0em}

\section{Introduction}
Dissipative Kerr solitons (DKS) in optical microresonators have emerged as robust, chip-scale, low-power-consumption frequency-comb sources \cite{kippenberg2018dissipative}. Research on DKS has ranged from fundamental studies of nonlinear-optical dynamics traceable to the Lugiato-Lefever equation (LLE) to a wide variety of applications that benefit from compact light sources. Studies on DKS nonlinear dynamics include breathing oscillations \cite{bao2016observation,lucas2017breathing} and their spontaneous synchronization \cite{cole2019subharmonic}, dark-pulse formation \cite{xue2015mode}, soliton crystallization \cite{cole2017soliton}, and exploration of new configurations for soliton generation with pulsed \cite{obrzud2017temporal} and chirped \cite{cole2018kerr} pump lasers.  Some applications of DKSs include direct frequency comb spectroscopy \cite{stern2018direct}, dual-comb spectroscopy \cite{suh2016microresonator,dutt2018chip}, distance measurements \cite{suh2018soliton,trocha2018ultrafast}, 
and optical telecommunications \cite{marin2017microresonator}.  Octave spanning solitons with dual dispersive waves (DWs) \cite{li2017stably,pfeiffer2017octave,briles2018interlocking} capable of being self-referenced \cite{jones2000carrier} provide a coherent link between the RF and optical domains, enabling optical frequency synthesis \cite{briles2018interlocking,spencer2018integrated} and chip-scale optical atomic clocks \cite{newman2019architecture}.

Stoichiometric silicon nitride (Si$_3$N$_4$) is a premier photonic material in the near-infrared, due to its high nonlinearity, large transparency window, and compatibility with lithographic fabrication techniques \cite{moss2013new}.  This latter feature allows for sophisticated group-velocity dispersion designs and integration with other photonic elements. Previous demonstrations of octave-spanning single-soliton frequency combs in Si$_3$N$_4$ \cite{li2017stably,briles2018interlocking,pfeiffer2017octave} have used high on-chip pump powers to overcome sub-optimal conditions of resonator group-velocity dispersion, waveguide-resonator coupling, and low intrinsic quality factor ($Q_\text{i}$) of the resonator. 
Lately, there have been numerous efforts to generate solitons with low pump power \cite{ji2017ultra,liu2018ultralow} and with chip-scale semiconductor lasers without an additional optical amplifier \cite{stern2018battery,pavlov2018narrow,raja2019electrically}. Despite these impressive photonic-integration developments, the bandwidth of these soliton frequency combs have been limited to less than 10 THz ($\approx 6.5 \%$ of an octave for solitons pumped at 1550 nm). Moreover, direct generation of octave-spanning soliton combs with a semiconductor laser is a critical step for future development of integrated-photonics frequency comb systems, but has remained an outstanding challenge. 

Here, we demonstrate generation of octave-spanning soliton frequency combs in both the 1550 nm and 1064 nm wavelength bands, with less than 40 mW of on-chip pump power, as well as direct soliton generation with a compact semiconductor laser in the 1064 nm band. By optimizing the waveguide-resonator coupling in high $Q_\text{i}$ and dispersion-engineered resonators with 1 THz free spectral range, the threshold power for Turing pattern generation is as low as $1.3 \pm 0.4$ mW.  Further, we efficiently convert a fraction of the pump laser power into a 1 THz repetition-frequency soliton frequency comb, which spans more than one octave and features dual-dispersive-wave power enhancements to assist $f$-2$f$ frequency comb self referencing. These are the elements that are required to construct a compact frequency-comb system based on integrated photonics.

\begin{figure*}[t!]
\centering
\includegraphics[width = 0.95\textwidth]{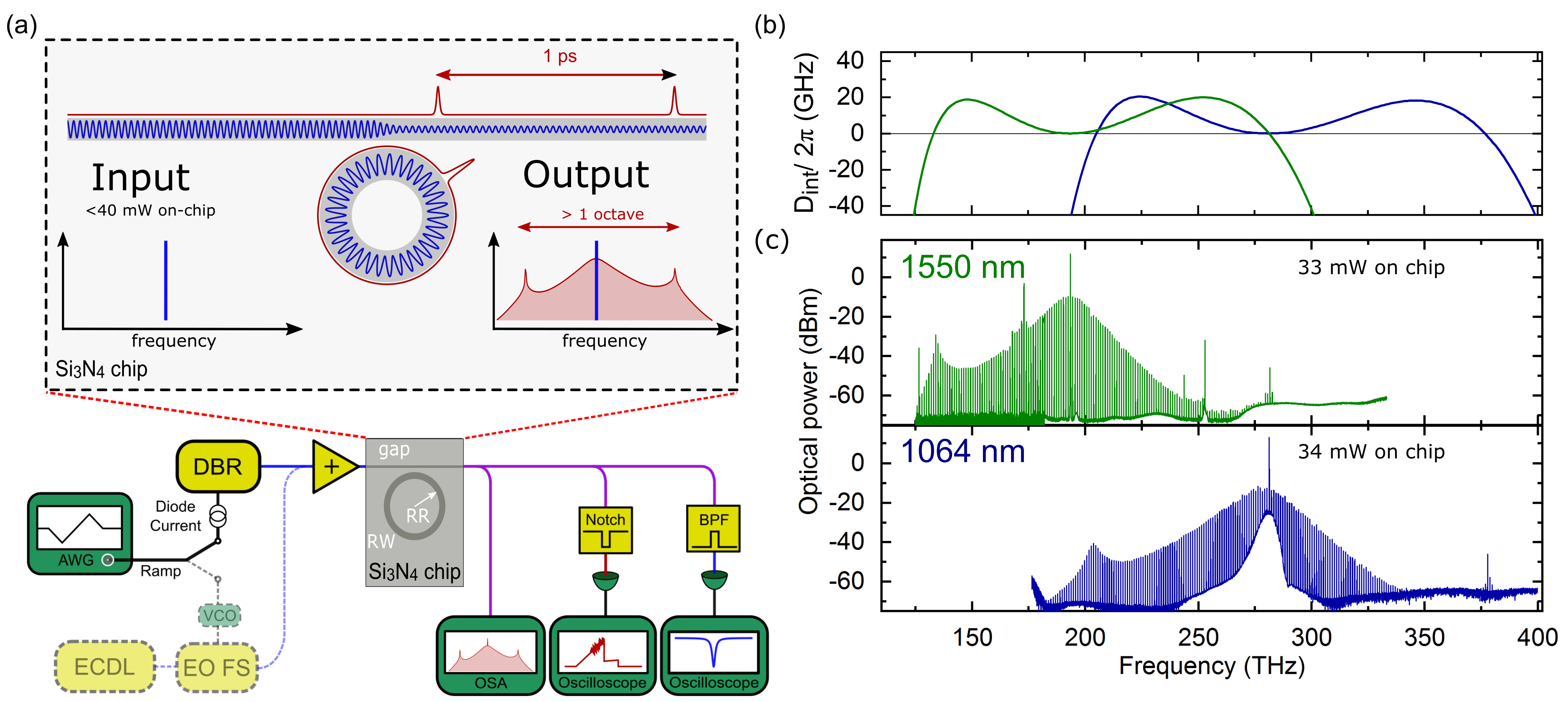}
\caption{\label{fig:Fig1}(a) Experimental setup for soliton generation in optimized Si$_3$N$_4$ resonators (bottom) and an enlargement highlighting key processes occurring on chip. AWG, arbitrary waveform generator; ECDL, external cavity diode laser; EO FS, electro optic frequency shifter; DBR, distributed Bragg reflector laser; OSA, optical spectrum analyzer; BPF, band pass filter. (b) Dispersion design and (c) octave spanning single soliton spectra obtained at low pump powers at pump wavelengths of 1550 nm (green) 1064 nm (blue). 
}
\end{figure*} 

This paper is organized into the following Sections.  Section \ref{sec:methods} describes the resonator dispersion design and the generalized experimental setup, including the diagnostic tools for reliable soliton generation. In Section \ref{sec:resonatorDESIGN} we discuss the optimization of the waveguide-resonator coupling rate for Kerr parametric oscillation threshold powers at the 1 mW level.  Section \ref{sec:FASTsweepPRINCIPLES} describes the principles of soliton generation in the limit of laser detuning control that is much faster than the resonator's thermal dynamics.  In Section \ref{sec:DWvsPOWER} we characterize the bandwidth-power scaling in our devices, including the dependence of dispersive-wave power on pump-resonator detuning.  In Sec. \ref{sec:DIODEpumping} we use the principles outlined in Sections \ref{sec:methods}-\ref{sec:DWvsPOWER} to enable direct diode laser pumping of octave bandwidth solitons at on-chip powers of 40 mW.  


\section{Methods
\label{sec:methods}}

Figure \ref{fig:Fig1}(a) outlines our experimental setup, which is composed of a continuous-wave (CW) pump laser, a SiN photonic chip, and control and characterization instruments. We utilize the Ligentec\footnote{Mention of specific companies or trade names is for scientific communication only, and does not constitute an endorsement by NIST.} silicon nitride foundry 
for wafer scale fabrication of $\approx$ 100 photonic chips, each containing a series of $\approx 75$ ring resonators and associated access waveguides with a ring coupling section and inverse taper structures at the chip edges for mode conversion to lensed fiber.  The CW laser that we frequency tune onto resonance energizes a soliton pulse train whose repetition frequency is derived primarily from the ring resonator free-spectral range. Incorporating numerous devices on one chip allows us to systematically vary the parameters that influence soliton generation and the frequency-comb spectrum, for example the ratio of the internal loss rate to the output coupling rate and the group-velocity dispersion (GVD or simply dispersion).




We design SiN ring resonators to produce octave-bandwidth soliton frequency combs, based on knowledge of the resonator material GVD \cite{luke2015broadband} and 2D axisymmetric finite-element simulations of the ring resonance frequencies (\emph{COMSOL Multiphysics}\footnotemark[\value{footnote}]). Since all the devices are fabricated on a single wafer, we target our designs around a silicon-nitride device layer thickness of 652 nm (780 nm) for 1064 nm (1550 nm) laser pumping and a rectangular cross section. We currently understand that the Ligentec foundry process we use produces approximately vertical sidewalls, which simplifies soliton-comb design. By varying the ring-resonator waveguide width (ring width, $RW$), we adjust the overall GVD profile to continuously change the bandwidth of the soliton comb from only 10's of nm to more than an octave \cite{briles2018interlocking,yu2019tuning}. The dominant effect is modification of the second-order GVD coefficient. However, higher-order dispersion is critical to broaden the comb spectrum through dispersive-wave (DW) generation. The conditions to realize a DW are: (1) near-equivalence of a resonator mode frequency and the corresponding comb mode, and (2) balance of four-wave mixing gain and resonator loss. The first condition is addressed in Fig. \ref{fig:Fig1}(b) which shows shows a simulation of the integrated GVD ($D_\text{int}/ 2\pi$) whose zero-crossings approximate the optical frequency of DWs in the soliton comb spectrum. Here, the parameter $RW=$ 810 nm ($RW=$ 1710 nm) is optimized for 1064 nm (1550 nm) wavelength pumping of an octave-bandwidth soliton comb. However, a prerequisite to utilizing high-resolution lithography to shape soliton frequency combs for applications is a reliable soliton generation technique.  

In this work, we generate solitons with high fidelity by agile control of the pump-laser frequency. We describe two methods, depending on the type of semiconductor laser used -- either a tabletop external-cavity diode laser (ECDL) or a compact, distributed-Bragg-reflector (DBR) diode laser; see the bottom of Fig. \ref{fig:Fig1}(a). The pump laser is amplified in either an ytterbium doped fiber amplifier (1064 nm) or an erbium doped fiber amplifier (1550 nm). The generic wavelength-dependent optical amplifier is represented in Fig. \ref{fig:Fig1}(a) as a yellow triangle with an inscribed plus sign. The amplifier overcomes system losses between the lasers and the on-chip ring resonators, arising primarily from mode conversion between optical fiber and the integrated waveguides. Both methods rely on fast laser frequency sweeps to mitigate destabilizing thermal transients; see Secs \ref{sec:FASTsweepPRINCIPLES} and  \ref{sec:DIODEpumping} for details.  

In the ECDL-based method \cite{briles2018interlocking,stone2018thermal,yu2019tuning}, we use a voltage-controlled oscillator (VCO) to drive an electro-optic frequency shifter (EO FS). The VCO sweeps the laser frequency by 5 GHz in a duration of 100 ns.  The components and system connections of the EO FS are shown with dashed lines in the lower branch of Fig. \ref{fig:Fig1}(a).  In the DBR diode laser method, the laser frequency is controlled by direct modulation of the diode current of a compact 1064 nm distributed Bragg reflector (DBR) laser from Photodigm\footnotemark[\value{footnote}] for frequency sweeps of $6.5$ GHz in 1.25 $\mu$s; Sec \ref{sec:DIODEpumping} provides a detailed description.  In both methods, the frequency ramp profiles are generated by an electronic arbitrary waveform generator (AWG).  

We diagnose soliton generation using several techniques.  Spectral domain analysis of stable solitons is performed by diverting a small fraction of the total light coupled off chip to an optical spectrum analyzer (OSA).  For time-domain analysis, we direct the majority of the remaining light to a notch filter centered at the pump wavelength to isolate the generated comb light from the unconverted pump light.  Observation of the photodetected filtered comb power on an oscilloscope during laser sweeps provides a critical diagnostic of soliton formation (see Fig. \ref{fig:Fig3}). 
High-bandwidth electronics ($> 150$ MHz) allow the boundary between the high-noise, modulation instability (MI) regime at negative (blue) laser detuning and the the low-noise, soliton regime at positive (red) detuning \cite{kippenberg2018dissipative} to be distinguished with high signal-to-noise; see bottom right of Fig. \ref{fig:Fig1}(a).

By use of the EO FS technique and the SiN ring resonators described above, we create the octave-bandwidth soliton combs displayed in Fig. \ref{fig:Fig1}(c). The DW wavelengths are in reasonable agreement with the $D_\text{int}$ approximation. Moreover, generating these soliton combs requires 34 mW of on-chip power.  The coupling losses at the facets are 3 dB and 6.5 dB for the 1550 nm and 1064 nm devices, respectively. Importantly, DWs offer high power per mode, comparable to that near the center frequency of the comb. However, realizing this condition requires efficient resonator outcoupling across $\sim200$ THz and operation of the pump laser at large red detuning; satisfying these conditions in optimally designed resonators is still a challenge.



\section{Optimizing external resonator coupling
\label{sec:resonatorDESIGN}}

\begin{figure}[b!]
\centering
\includegraphics[width = 0.95\linewidth]{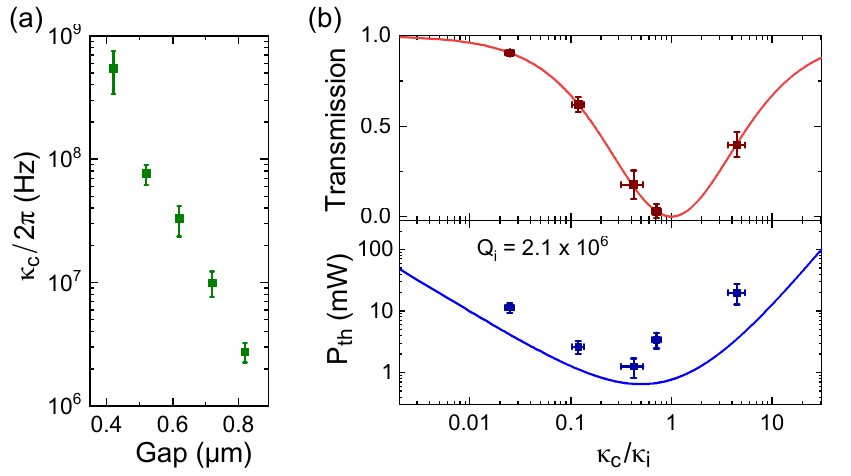}
\caption{\label{fig:Fig2} Optimization of waveguide-resonator coupling rate $\kappa_{\text{c}}/2\pi$ for a 1064 nm design.  (a)  The measured $\kappa_\text{c}/2\pi$ as a function of nominal gap distance averaged over measurements from $280-294$ THz.  (b) Measured on-resonance pump (top) and four-wave mixing threshold power as a function of the coupling parameter $\kappa_\text{c}/\kappa_\text{i}$ over the bandwidth of our optical amplifier ($280-284$ THz).  
}
\end{figure}

To achieve the lowest power soliton microcomb operation, we carefully optimize the external coupling rate of the resonator to the on-chip waveguide. Here we review the analysis of this problem, and we present data to evaluate the results of our optimization. The Lugiato-Lefever equation (LLE) \cite{lugiato2018lugiato,coen2013modeling} provides an accurate model for soliton behavior in our resonators. In particular, we characterize the pump laser power ($P_\text{in}$) normalized to the threshold power for parametric oscillation ($P_\text{th}$) as $F^2 = P_\text{in}/P_\text{th}$. Resonator loading is described in terms of the coupling parameter $K = \kappa_\text{c}/\kappa_\text{i} = Q_\text{i}/Q_\text{c}$, where $\kappa_\text{i}$ is the intrinsic loss rate of the resonator and $\kappa_\text{c}$ is the coupling rate.  It is also important to characterize the pump laser frequency detuning from the cavity resonance $\Delta$, which we normalize to half the resonator total loss rate $\kappa_{\text{tot}} = \kappa_{\text{i}} + \kappa_{\text{c}}$ so that $\Delta = \frac{2\pi \left( \nu_0 - \nu_{\text{p}} \right)}{\kappa_{tot}/2}$.  Here, $\nu_0$ and $\nu_{\text{p}}$ are the optical frequency of the resonator and pump modes respectively.  The effects of detuning are discussed in more detail in Sections \ref{sec:FASTsweepPRINCIPLES} and \ref{sec:DWvsPOWER}.  These quantities are related to the associated quality factors $Q_\text{c}$ and $Q_\text{i}$ by $\kappa_\text{c}/2\pi = \nu_0/Q_\text{c}$ and $\kappa_\text{i}/2\pi = \nu_0/Q_\text{i}$ respectively.  The threshold power for parametric oscillation \cite{kippenberg2004kerr,li2012low} is
\begin{equation}
    P_\text{th} = \frac{\pi}{8} \frac{n}{n_2}\frac{\nu_0}{\nu_{\text{FSR}}}\frac{A_\text{eff}}{Q_\text{i}^2} \times \frac{\left( 1+K\right)^3}{K},
    \label{Eq:THRESHOLDpower}
\end{equation}
where $n$ is the refractive index, $n_2$ is the nonlinear refractive index, $\nu_\text{FSR}$ is the resonator free spectral range (FSR), $A_\text{eff} \approx 0.5$ $\mu$m$^2$ is the effective mode-area of the resonator. One may optimize the resonator material selection and other properties to reduce $P_\text{th}$. Interestingly, Eq. (1) indicates that the absolute minimum threshold occurs for a setting of $K=0.5$.


Figure \ref{fig:Fig2} presents our experimental data and analysis of threshold power in devices designed for use with a 1064 nm laser. Operationally, we characterize the resonator $Q_\text{i}$ and $Q_\text{c}$ values with measurements of a device's transmitted $P_\text{in}$.  The minimum transmission on resonance $T$ is related to $K$ by the relation $T=\left(\frac{1-K}{1+K} \right)^2$.  In Fig. \ref{fig:Fig2}(a), we explore how the external coupling rate $\kappa_\text{c}/2\pi$ changes with the geometrical gap size between the resonator and the waveguide. These devices use a straight waveguide coupler with a single point of closest approach to the resonator. (The external coupling rate is also controlled by the geometry of the coupling region.) For these measurements, we use different device samples on the same chip with a programmed variation in the gap size, and we perform these measurements at low power ($\approx$ 50  $\mu$W) so that thermal bistability and optical non-linearity are negligible. We measure five devices with nominally identical RW's (0.85 $\mu$m), access waveguide widths (0.51 $\mu$m) and film thickness (652 nm), but with waveguide-resonator gaps ranging from $420 - 820$ nm in 100 nm steps. Moreover, we measure all the resonances in the optical-frequency range 280 to 294 THz. The error bars correspond to the standard deviation of these measurements. As expected, $\kappa_\text{c}$ decreases nearly exponentially with gap distance.


With our characterization of $K$, we present measurements of the resonator transmission $T(K)$ (Fig. \ref{fig:Fig2}(b), top) and the parametric oscillation threshold power $P_\text{th}(K)$ (Fig. \ref{fig:Fig2}(b), bottom) alongside the expected theoretical behavior (solid lines). We characterize $P_\text{th}(K)$ by the in-waveguide pump-laser power required to generate signal and idler sidebands when the laser frequency is tuned to resonance. Here, the measurements were performed over the optical-frequency range $280-284$ THz, and the error bars indicate one standard deviation. The transmission measurements demonstrate under-, critical- and over-coupling of the resonator depending on the gap. The measurements of $P_\text{th}$ demonstrate the most power efficient parametric oscillation with slight under-coupling, i.e., $\kappa_\text{c} = 0.5 \, \kappa_\text{i}$. These behaviors match our expectations from Eqn. (\ref{Eq:THRESHOLDpower}), as indicated by the solid-line model predictions in Fig. \ref{fig:Fig2}(b), which we obtain based on the measured $Q_\text{i}$ and no other free parameters. Indeed, the threshold power for oscillation with these devices is approximately 1 mW.

\section{Soliton generation with fast detuning control}
\label{sec:FASTsweepPRINCIPLES}

The primary benefit of the EO FS technique is increased soliton stability resulting from the very high sweep rates that can be obtained.  These sweeps are well suited to the unique challenge of fast thermal dynamics in compact SiN resonators \cite{li2017stably,brasch2016bringing,arbabi2012dynamics} because they enable detuning control at the primary thermal relaxation time scale ($\tau_{\text{th}} \approx 100-200$ ns) of the ring resonator.  Hence, an EO FS can be programmed to mitigate thermal instabilities associated with the transition between the MI and soliton regimes. In this section we give a detailed procedure for using an EO FS to optimize soliton formation.  The ideas presented here directly extend to the diode current modulation method presented in Sec. \ref{sec:DIODEpumping}.

According to the LLE, DKS properties are determined by the resonator nonlinearity, dispersion and loss, as well as two external parameters; the pump laser power and the relative pump-resonator detuning \cite{herr2014temporal,yi2015soliton}. Importantly, there is a pump-power dependent detuning existence range over which the soliton can stably propagate.  It is bound on one end by the boundary between the MI regime and soliton regime at $\Delta_{\text{min}} \geq \sqrt{3}$. In the absence of higher order dispersion, the maximum accessible detuning is given by $\Delta_{\text{max}}= \pi^2F^2/8$.  Within the soliton existence range, theoretical scaling laws \cite{coen2013universal} predict that the soliton peak power and bandwidth monotonically increase with detuning \cite{yi2015soliton, yi2016active}.  We note that the scaling law for soliton power has recently been extended to include Raman effects \cite{li2018universal}.  Although an analytic estimate for $\Delta_{\text{max}}$ is unavailable in the case of strong higher order dispersion, our numerical simulations of the LLE indicate that $\Delta_\text{max} = \pi^2 F^2/8$ remains approximately correct.  

Here we are interested in the detuning dynamics caused by programmed laser frequency sweeps as well as thermal relaxation of the resonator.  Accordingly, we modify the expression for the detuning to explicitly include dependence on temperature $T$,

\begin{equation}
\Delta (T) = \frac{2\pi \left[ \nu_0 (T) - \nu_{\text{p}} \right]}{\kappa_{tot}/2}
\label{Eq:tempDEPENDENTdetuning}
\end{equation}



\begin{figure*}[htbp]
  \centering
\includegraphics[width = 0.95\textwidth]{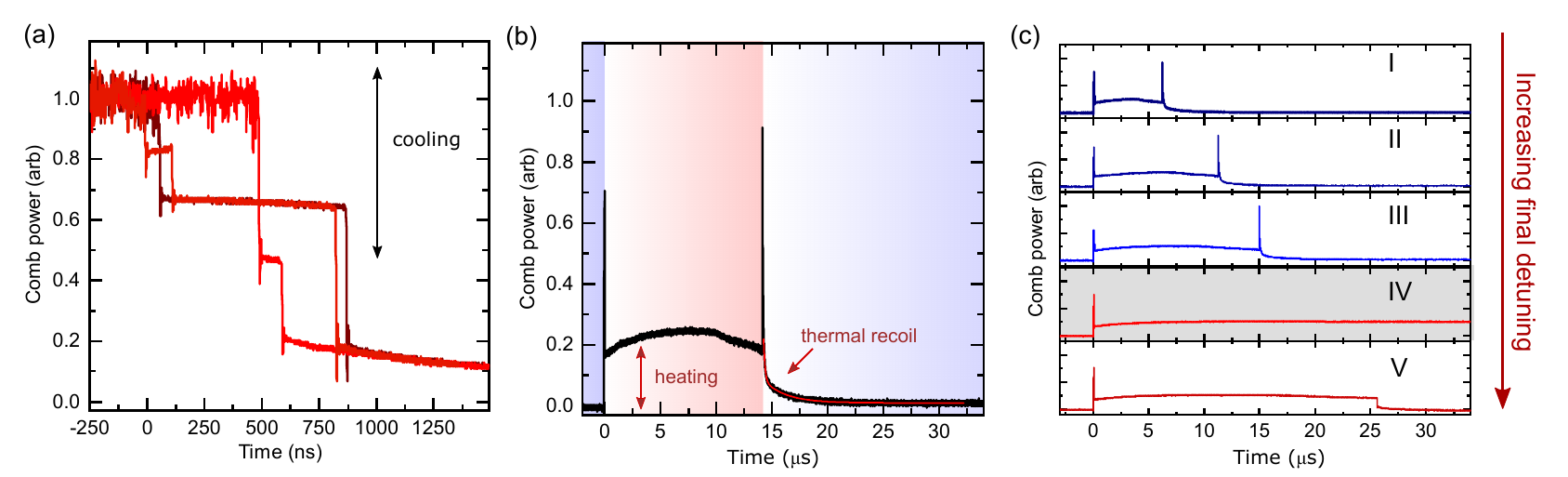}
\caption{\label{fig:Fig3} Soliton thermal dynamics and optimization of fast laser frequency sweeps with an EO FS pump source.  (a) Slow sweep regime: an adiabatic sweep across the cavity resonance results cooling and soliton step durations that are limited by the resonator's thermal relaxation rate.  The soliton is lost when $\Delta > \Delta_{\text{max}}$ (i.e.: too red).  (b) Fast sweep regime: A sweep that accesses the soliton state in a time shorter than $\tau_{\text{th}}$ results in delayed heating.  The soliton state is lost when $\Delta < \Delta_{\text{min}}$ (i.e.: too blue).  The step durations are typically longer than in the adiabatic case due to the smaller magnitude $\Delta T$.  (c) Optimizing soliton generation with fast sweeps using the principles of (a) and (b).  Increasing the final detuning results in an increased step duration for panels I-III followed by eventual loss of the soliton state at small blue detunings when $\Delta < \Delta_{\text{min}}$ as in (b).  A stable soliton state is reached in panel IV.  Further increases result in loss of the soliton state at large red detunings, $\Delta > \Delta_{\text{max}}$ as in (a).  Red (blue) traces correspond to cases where the final state is effectively red (blue) detuned. }
\end{figure*} 

The benefit of fast frequency sweeps is best illustrated by studying soliton thermal dynamics in the limit of slow detuning control.  Figure \ref{fig:Fig3}(a) shows the evolution of soliton power as the pump laser (initially blue detuned) is scanned over a range of $>30$ GHz at a rate $\approx 500$ GHz/s.  Although the data shown in Fig. \ref{fig:Fig3} is obtained with air-clad devices having a SiN thickness of 615 nm and RW = 1788 nm \cite{briles2018interlocking}, the results apply to oxide clade resonators as well.  The transition from the MI state to the soliton states with either $N=3,2,1$ pulses is accompanied by a precipitous drop in intracavity power resulting in significant cooling.  As a result, $\nu_0$ blue shifts towards its `cold-cavity' frequency at the resonator's thermal relaxation rate.  Once $\Delta (T)$ exceeds the existence range (i.e.: too red, $\Delta > \Delta_{\text{max}}$), the soliton is lost over a duration comparable to the photon lifetime (a few ns).  The duration of the soliton steps in this regime of adiabatic detuning control is a complicated function of the change in intracavity power and temperature \cite{li2017stably} as well as the extent of the pump-power-dependent soliton existence range (see Sec. \ref{sec:DWvsPOWER}), but are generally the same order of magnitude as $\tau_{\text{th}}$ \cite{li2017stably,brasch2016bringing,arbabi2012dynamics}. 
 The absence of step durations $>1 \mu$s even when the scan speed is reduced by over an order of magnitude indicate that thermal relaxation is the dominant mechanism that determines the step duration in the adiabatic regime.

Figure \ref{fig:Fig3}(b) examines soliton dynamics in the opposite regime, where the pump laser is tuned faster than the thermal dynamics. The pump laser starts blue detuned and its frequency is decreased by 5 GHz in under 100 ns, reaching the effectively red detuned single soliton state at $\approx 0$ ns. Because the soliton state is accessed so quickly, the resonator is still in approximate thermal equilibrium with the ambient environment immediatley after generating the soliton.  The sudden increase in intracavity power results in delayed `thermal charging', which red-shifts $\nu_0$ and eventually causes the soliton to be lost at small detunings when $\Delta < \Delta_{\text{min}}$ ($\approx 14 \ \mu$s in Fig. \ref{fig:Fig3}(b)).  Since the change in intracavity power between the cold-cavity and the single-soliton state is typically smaller than the change between the MI state and the single-soliton state, the corresponding temperature changes are typically smaller for the fast sweeping method compared to adiabatic sweeps.  This often results in longer soliton steps as shown in Fig. \ref{fig:Fig3}(b) and is a principle benefit of this method.  After loss of the soliton state, the comb traces back over the blue detuned MI states. A double exponential fit to the decaying tails on the blue detuned side, gives time constants of $ \tau_\text{th}^{(1)}\approx 100$ ns and $\tau_\text{th}^{(2)}  \approx 1.5 \ \mu$s.  The faster of these is attributed to thermalization of the $\approx 0.5 \ \mu \text{m}^2$ transverse mode area and the slower with the thermalization of the resonator as a whole \cite{ilchenko1992thermal}.  A third timescale on the order of $100 \ \mu\text{s}$ - 1 ms is not evident in Fig. \ref{fig:Fig3}(b), but it is attributed to thermalization of the entire chip.  The third time scale and its associated temperature changes can be altered somewhat by improving the thermal conductivity between the chip and its mount.

Figure \ref{fig:Fig3}(c) outlines how the general behavior illustrated in Fig. \ref{fig:Fig3}(a) and Fig. \ref{fig:Fig3}(b) are used to optimize the laser frequency ramp for successful generation of solitons.  To stabilize the soliton state, the final value of the detuning is progressively increased to chase the thermally red-shifting resonance.  Each of the five panels in Fig. \ref{fig:Fig3}(c) corresponds to a different final detuning, which increases monotonically from panel I to panel V.  In panels I-IV, the increase in final detuning leads to an increased duration of the soliton state until a stable soliton is generated that can circulate indefinitely within the cavity (panel IV).  Increasing the detuning past the existence range ($\Delta > \Delta_{\text{max}}$) results in the soliton being lost within a few ns without retracing the blue detuned MI regime (panel V) similar to the traces in Fig. \ref{fig:Fig3}(a).  

In samples with large thermo-optic shifts, it may be necessary to include $2^{nd}$ and $3^{rd}$ order ramps to address the higher-order thermal relaxation time scales.  Since these additional ramps are active while the resonator temperature is relaxing, one must consider the dynamics of both $\nu_0$ and $\nu_p$ in Eq. \ref{Eq:tempDEPENDENTdetuning}.  The additional ramps should be optimized using the slope of the detuning-dependent soliton power oscilloscope trace.  Because of the positive correlation between soliton power and detuning \cite{yi2015soliton,li2018universal}, a change in soliton power over time indicates a commensurate change in $\Delta$.  In particular, if the soliton power is increasing (decreasing) during the additional ramp segments it suggests that the applied ramp rate is too fast (slow) compared to the relevant thermal relaxation.  The rate should be adjusted until a flat soliton power trace is obtained.  The final detunings of these ramps should be adjusted according the principles outlined in Fig. \ref{fig:Fig3}(c).

\section{Optimizing the soliton frequency comb with laser detuning
\label{sec:DWvsPOWER}}

Following initiation of the soliton frequency comb with either frequency-agile scanning technique, we perform a fine adjustment of the pump-cavity detuning that is designed to increase the soliton comb bandwidth, and in particular, increase the intensity of both the short-wavelength and long-wavelength dispersive waves (SDW and LDW respectively) which is important for self-referencing.    In this section, we demonstrate the results of this fine adjustment.

 In the absence of interference from the phase dependent CW background and the DW's \cite{skryabin2017self}, the DW intensities generally increase monotonically with detuning, reaching the maximum just before the soliton is lost at the edge of the soliton existence range.  As a result, for a given pump power the most intense DW's are found near $\Delta_\text{max}$.  To increase the DW intensity further, larger $F^2$ values are required to reach correspondingly larger  values of $\Delta_\text{max}$.

\begin{figure}[htbp]
\centering
\includegraphics[width = 0.95\linewidth]{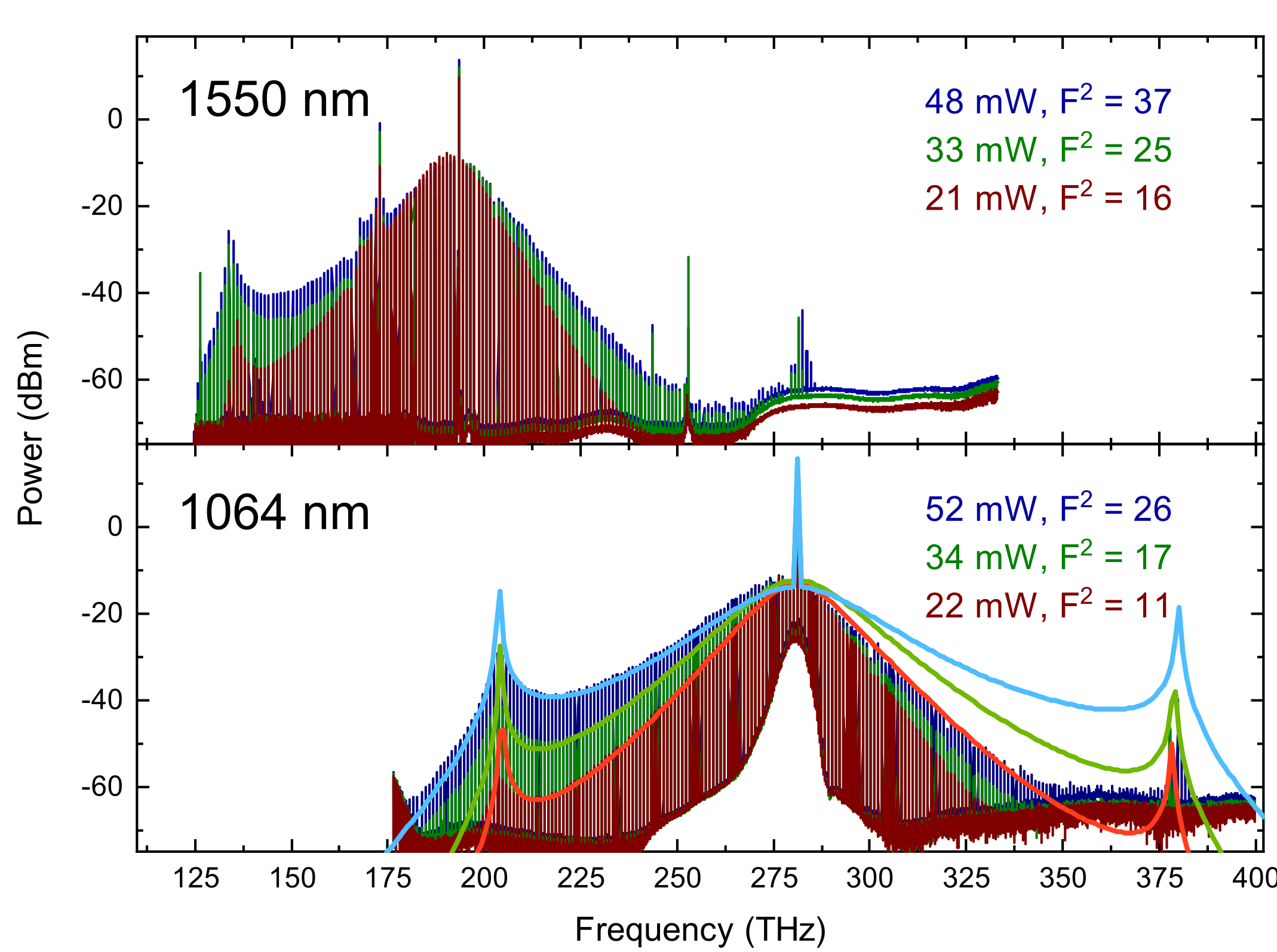}
\caption{\label{fig:Fig4} Soliton spectrum for different settings of pump-laser power. The top (bottom) panel displays data with a 1550 nm (1064 nm) laser. For the 1064 nm data, the spectral envelopes from LLE simulations are shown as solid lines.  The large discrepancy in spectral power at high optical frequencies is attributed to poor coupling conversion from the ring resonator to the on-chip coupling waveguide.
%
}
\end{figure}

Figure \ref{fig:Fig4} investigates how the solitons shown in Fig. \ref{fig:Fig1} change with pump power and pump-resonator detuning. Here the EO FS technique was used in both the 1550 nm and 1064 nm wavelength bands but the results are general and independent of the soliton generation method.  We note that amplified spontaneous emission is not removed with a bandpass filter before the resonator for the 1064 nm spectra (bottom panel).   Once solitons are obtained, the laser frequency is decreased to grow the soliton bandwidth, with the broadest spectra occurring near the experimentally determined value of $\Delta_\text{max}$.  All spectra in Fig. \ref{fig:Fig4} are recorded at this condition.  Solitons are initially generated with approximately 20 mW of pump power (red traces) and each shows relatively weak LDWs and no measureable SDW.  Significant SDWs are not obtained until the pump power is increased by $50\%$ to 33-34 mW (green traces) corresponding to $F^2 = 25$ and $17$ for the 1550 nm 1064 nm cases, respectively.  For both cases the LDW increases by $17$ dB and the SDW emerges at -46 dBm. Another $50\ \%$ increase in the pump power results in smaller increases in DW intensity. Numerical LLE simulations for the 1064 nm case show similar dependence of DW intensities on pump power and detuning.  Simulated soliton envelopes near $\Delta_{\text{max}}$ are shown in the bottom panel of Fig. \ref{fig:Fig4} for each power under consideration.  The general behavior reflected in the experimental and simulation results has important practical consequences for self-referencing applications.  Namely, it suggests that much higher pump powers are required to generate significant DW's than are required for simple soliton generation.   We also note that the large discrepancy in the absolute SDW power between the experiment and simulation is in agreement with the expected poor performance of straight waveguide couplers at short wavelengths \cite{moille2019broadband}.



\section{Direct Diode-Laser Pumping of octave-bandwidth soliton combs
\label{sec:DIODEpumping}}



\begin{figure}[htbp]
    \centering
    \includegraphics[width = 0.95\linewidth]{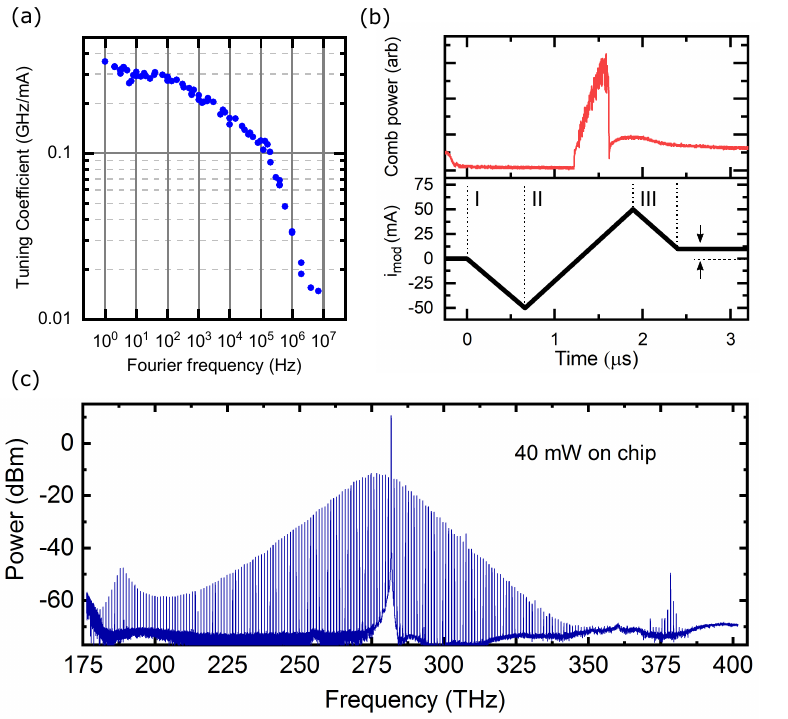}
    \caption{Soliton generation with direct modulation of laser diode current. (a) Measured frequency modulation response of DBR laser. (b) Photodetected comb power during soliton generation (top) and corresponding applied 3-segment laser current ramp (bottom).  (c) Spectrum of single soliton obtained with 40 mW of on-chip pump power.}
    \label{fig:Fig5}
\end{figure}

Given our detailed understanding of the EO FS technique that enables reliable generation of stable octave-span solitons (Sec. \ref{sec:FASTsweepPRINCIPLES}), we adapt this approach for use with more compact and integration-ready semiconductor lasers. This offers the potential of simplified and lower power consumption comb system. Figure \ref{fig:Fig5} shows our characterization of diode lasers for this purpose and our observation of octave-span soliton combs.  

The primary challenge that arises in adapting the EO-FS technique to direct diode-current modulation is duplicating the simultaneously large range and short duration frequency sweep. Current modulation of diode lasers is complicated by the presence of two distinct mechanisms responsible for changing the laser frequency: (1) carrier density effects, and (2) direct Joule heating of the diode junction \cite{coldren2012diode}.  At low Fourier frequencies the modulation response is dominated by thermal effects where an increase in current heats the diode and leads to a decrease of the laser frequency.  Beyond the thermal cutoff frequency $f_{\text{c}}$, the response is dominated by carrier density effects, where an increase in diode current leads to an increase of the laser frequency. The value of $f_{\text{c}}$ depends on the geometry and composition of the laser cavity but is typically in the range of 10 kHz -  1 MHz.  However, the opposite phase of the thermal and carrier density tuning coefficients complicates optimization of the ramp waveform for soliton generation. Therefore, we avoid altogether the carrier density regime.  In this case, achieving a high-bandwidth frequency modulation requires a small volume laser cavity that offers a fast thermal relaxation time.   



Figure \ref{fig:Fig5} (a) shows the measured frequency-modulation response of a short-cavity DBR laser operating at $1064\pm 0.5$ nm.  The free-space laser output passes through an optical isolator before coupling to fiber. The magnitude response is measured as a function of Fourier frequency between $10^0-10^7$ Hz with a calibrated, 500 MHz Mach-Zehnder interferometer.  The low-frequency tuning coefficient is $0.35$ GHz/mA and falls to $0.1$ GHz/mA by 100 kHz.  

Based on the modulation response of the DBR laser, we design a laser frequency ramp sequence for soliton generation using the results of Sec. \ref{sec:FASTsweepPRINCIPLES}. The optimized ramp profile is an asymmetrical triangle waveform, consisting of three segments; see bottom panel of Fig. \ref{fig:Fig5}(b). This modulation current $i_{\text{mod}}$ is added to a DC bias current of several hundred mA.  Initially, the laser is set to a positive (red) frequency detuning, and minimal pump power is coupled to the resonator.  In ramp segment I, the current is decreased by 50 mA, which tunes the laser frequency above the cavity resonance (blue-detuned).  In segment II the current is increased by 100 mA, decreasing the frequency by 6.5 GHz in 1.25 $\mu$s. This ramp segment generates a modulation instability pattern in the resonator that seeds the generation of a single soliton; see top panel of Fig. \ref{fig:Fig5}(b).

Once the soliton is generated at the end of segment II, we apply a final segment (III) to ensure that the soliton is stable.  In contrast to segment II, where ramp parameters are chosen to minimize thermal transients in the SiN resonator, the parameters of segment III  are chosen to minimize thermal transients in the DBR laser cavity.  The finite bandwidth of the current modulation response leads to discrepancies between the applied ramp profile and the actual change in laser frequency for operation at the edge of the modulation bandwidth as done here.  These effects are most important at the end of the ramp sequence where the ramp's high Fourier-frequency components associated with the transition to a DC level are filtered out, causing the laser frequency to drift after the applied ramp has ended.  Because these drifts evolve at the DBR's natural thermal time scale which is significantly longer than the SiN resonator's $\tau_{\text{th}}$, they are incompatible with soliton stabilization.  The magnitude of the drift is increased if the final current value is different from the initial value at start of ramp segment I due to a change in the DBR cavity temperature.  We use a final current decrease in segment III to effectively counteract this drift.  The optimum magnitude of segment III is a balance of maintaining the soliton existence range and keeping the laser in approximate thermal equilibrium; see arrows in the bottom panel of Fig. \ref{fig:Fig5}(b).  If the magnitude of the current decrease is too small, the slow uncompensated drift in laser frequency causes the soliton to be lost at positive (red) detunings beyond $\Delta_{\text{max}}$.  If the current is decreased too far, the range of successful initial laser frequencies and pump powers is impractically small and the soliton will generally transition back to the MI regime at negative (blue) detunings.    Note that the reversed ramp direction in segment III is reflected in the local maximum of the soliton power shown in Fig. \ref{fig:Fig5} (b) and arises from the detuning dependence of the soliton power.








Figure \ref{fig:Fig5}(c) shows the spectrum of a soliton generated in the same resonator device as in the bottom panel of Fig. \ref{fig:Fig4} using the frequency-swept DBR laser technique. Here, we utilize only 40 mW of on-chip power to generate and stabilize this soliton. Moreover, following stabilization of the soliton, we utilize the technique described in Section \ref{sec:DWvsPOWER} to optimize the soliton DWs. With the DBR laser, we increase the temperature of the laser package to reach the condition of highest intensity DW.



\section{Conclusion}

We have reported ultralow-power generation of octave-bandwidth, single-soliton frequency combs in integrated SiN photonic circuits. We achieve the lowest possible on-chip laser power consumption by optimizing the external coupling regime. Moreover, we describe reliable laser-frequency control techniques to generate soliton combs in an arbitrary SiN ring resonator. This is a key to realizing deterministic control of the soltion comb spectrum for applications like $f-2f$ self-referencing. Matching the optical mode profile of the chip facets to the DBR laser would allow direct pumping without an optical fiber amplifier. The principles of the soliton generation technique outlined here should be widely applicable to other tunable narrow linewidth semiconductor lasers \cite{tran2019tutorial} included distributed feedback diode lasers (DFBs) and extended DBR lasers integrated on silicon \cite{huang2019high}. 


\section{Acknowledgements}

We thank Cort Johnson and Christine Wang at Draper Laboratories for loan of the DBR laser used in these experiments, and Jizhao Zhang and Pablo Acedo for valuable comments on the manuscript. This project was funded by the DARPA ACES and DODOS programs and NIST. This work is a contribution of the U.S. government and is not subject to copyright.   

\bibliography{brilesBIBv16}

\end{document}